\documentclass[doublecol]{epl2} 
\usepackage{amsmath}
\usepackage{amsfonts}
\usepackage{verbatim} 
\usepackage[switch]{lineno}
\title{Thermodynamic uncertainty for run-and-tumble type processes}
\shorttitle{Thermodynamic uncertainty for run-and-tumble type processes} %Insert here a short version of the title if it exceeds 70 characters

\author{Mayank Shreshtha\inst{1} \and Rosemary J. Harris\inst{1}}
\shortauthor{Mayank Shreshtha and Rosemary J. Harris}

\institute{                    
  \inst{1} School of Mathematical Sciences, Queen Mary University of London, London, E1 4NS, UK \\
 }

\abstract{Thermodynamic uncertainty relations have emerged as universal bounds on current fluctuations in non-equilibrium systems. Here we derive a new bound for a particular class of run-and-tumble type processes using the mathematical framework of renewal-reward theory which can be applied to both Markovian and non-Markovian systems. We demonstrate the results for selected single-particle models as well as a variant of the asymmetric simple exclusion process with collective tumbles. Our bound is relatively tight for a broad parameter regime and only requires knowledge of the statistics of run lengths and the mean entropy production rate of tumbles.}

\begin{document}

\maketitle

\section{Introduction} 
\begin{figure}
\onefigure[scale = 0.25]{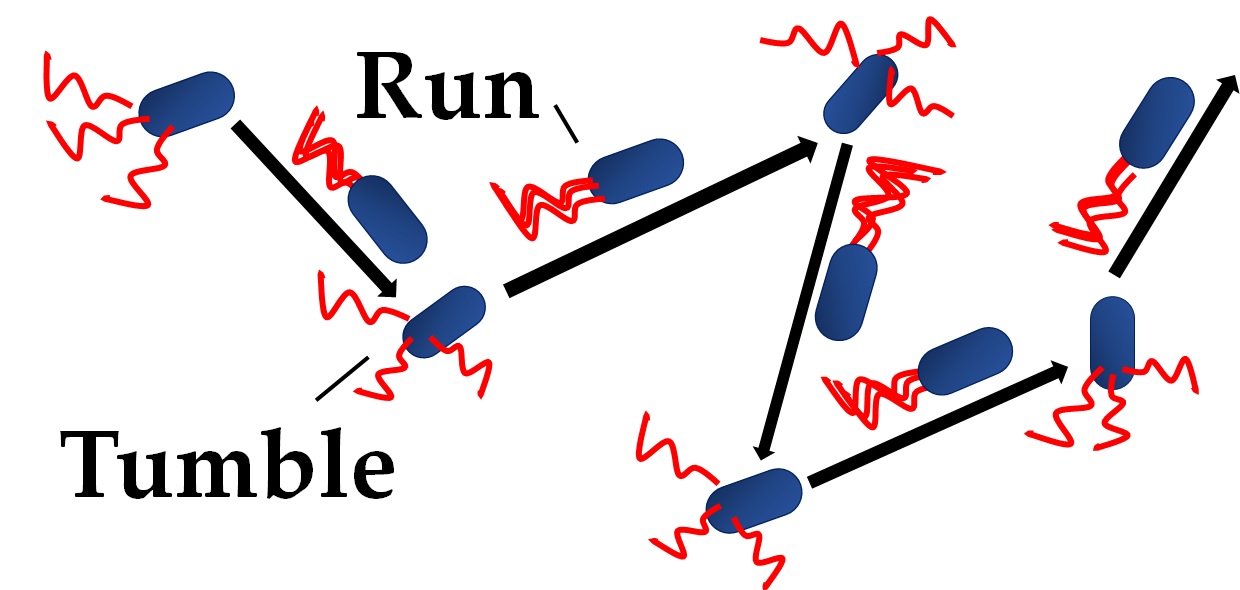}
\caption{ Typical bacteria locomotion: aligned bundle of hair-like projections (flagella) causes directed propulsion (runs) whereas spread-out bundle results in random reorientation (tumbles). }
\label{fig1}
\end{figure}

Recently, there has been a surge of interest in the study of thermodynamic uncertainty relations (TURs)~\cite{i4,i8} which quantify the universal trade-off between current (\textit{e.g., }velocity, particle-flux), its statistical fluctuations and entropy production. The vast majority of work on such TURs has been for time-homogeneous Markovian systems~\cite{i5, tur11, i9, tur12, tur8, i7, tur5, tur2, tur6, tur17,  tur3, tur18, tur4, tur10} although recent forays beyond this include discussions of periodically driven systems~\cite{tur9, tur15, tur16}, semi-Markov processes~\cite{tur19} and time-delayed Langevin dynamics~\cite{tur13, tur14}. Here, we consider current uncertainty in a general class of processes where random dynamics (runs) are punctuated by stochastic resets of preferred direction (tumbles). A particular example is the eponymous run-and-tumble process (see Fig.~\ref{fig1}) which provides a standard paradigm for bacterial motility~\cite{RT1, RT2, RT3} and is also used in modelling search strategies and various other systems (see~\cite{RT5} and references therein).

The presence of non-vanishing currents is an important characteristic of non-equilibrium stochastic systems. The time-integrated current $J$ typically follows a large-deviation principle~\cite{LDT1, LDT2} which implies that the cumulants scale with time $t$. In particular, for TURs, we are interested in the scaled mean and scaled variance:
\begin{equation}
\bar{j} = \lim_{t \rightarrow \infty} \frac{E[J]}{t}\ ,\qquad{\sigma_j}^2 = \lim_{t \rightarrow \infty} \frac{\text{Var [$J$]}}{t}.
\label{1}
\end{equation}

Uncertainty relations were originally given for continuous-time Markov processes but here we focus on a discrete-time version proposed by Proesmans and Van den Broeck~\cite{i7} and referred to as the ``PV bound"~\cite{tur18}:
\begin{equation}
\frac{\bar{j}^2}{{\sigma_j}^2} \leq \frac{1}{2 \Delta t} \left(e^{\bar{s}_{\text{tot}}}-1 \right).
\label{PV}
\end{equation}This inequality provides a constraint on the (reciprocal) uncertainty ${\bar{j}^2}/{\sigma_j}^2$ of any current $J$ in terms of the mean total entropy production rate $\bar{s}_{\text{tot}}$ of the process; $\Delta t$ is the time step which we can set to 1 without loss of generality.

In this letter, we derive another bound for the specific class of one-dimensional run-and-tumble models which involves  the mean entropy production rate $\bar{s}_X$ of an auxiliary process related only to the tumbles. This new bound takes the form
\begin{eqnarray}
 \frac{\bar{j}^2}{{\sigma_j}^2} \leq \left(\frac{(E[N-1])^2}{E[N] E[(N-1)^2]} \right) \times  \frac{1}{2} \left(e^{\bar{s}_X} -1 \right),
 \label{introRT}
 \end{eqnarray}where the prefactor contains  run-length statistics; more precisely, the random variable $N$ is the time between successive tumbles. Our derivation is based on the framework of renewal-reward theory (RRT)~\cite{RRTmain1}. RRT has previously been applied in modelling inventories, queueing and reliability~\cite{RRTbook} as well as biological systems (\textit{e.g.,} molecular motors~\cite{RRT1, RRT2} and stem-cell differentiation~\cite{RRT3}). Below, we expound in detail the connection between RRT and the integrated current in run-and-tumble type processes.

Significantly, our analysis shows that for many parameters, our bound~\eqref{introRT} is tighter than a naive application of the PV bound~\eqref{PV} in extended state space. We illustrate the new bound for single particles with geometric and non-geometric run lengths, as well as a many-particle exclusion process with collective reset. 
\section{Run-and-tumble random walk model}
We introduce here a simple one-dimensional run-and-tumble (RT) process in discrete space and discrete time which will serve as a toy model for the analysis of the following sections. In the spirit of the introduction, the process consists of alternating runs (biased random walks) and tumbles (which set the bias).  We assume the process always starts with a tumble: at $t = 0$, the preferred direction is set to be right (positive) with probability $p$ and left (negative) with probability $q = 1-p$. For all subsequent time steps $t =1, 2, 3,\ldots, T$, the particle tumbles with probability $f$ (resetting the preferred direction) and runs with probability $1-f$.  A run step consists of the random walker moving ``forwards'' (in the preferred direction set by the last tumble) with probability $p'$ and ``backwards" with probability $q' = 1-p'$. Note that for the special case $p'=1$,  our model reduces to a type of persistent random walk where the particle only changes direction when it tumbles. The observable of interest in this study is the time-integrated particle current $J(t)$, defined here as the net difference between the number of right and left steps up to time $t$. Since a tumble step serves only to set the  preferred direction, there is no current increment due to the tumble.

The duration of a combined tumble-and-run event (the time elapsed between successive tumbles) is clearly a random variable taking values $n = 1,2,3, \ldots$. Here the tumble occupies one time step and the run has length $n-1$; the case $n=1$ corresponds to tumbles at consecutive time steps. For the above-described dynamics, the $n$'s are drawn from a geometric distribution with parameter $f$. This simple process is Markovian on the extended state space of the position and preferred direction; by construction these variables are even under time-reversal.  The fluctuations of $J(t)$ in the long-time limit are encoded in the so-called scaled cumulant generating function (SCGF) \cite{LDT1}, $\phi(s) = \lim_{t \rightarrow \infty} (1/t) \ln E [e^{sJ}]$ which can be obtained via an eigenvalue problem in the extended state space, or within a reset framework~\cite{r1} as outlined in the Appendix. The SCGF yields the scaled cumulants of the current:
\begin{align}
\overline{j} = \phi'(0) &=  (p-q)(p'-q')(1-f), \label{5A}\\ 
{\sigma_j}^2 = \phi''(0)  &= (1-f) [f (p'-q')^2 +  4p'q'] \nonumber\\
& \phantom{=}~+\frac{(1-f)^2 (f+2)}{f}\left [4pq(p'-q')^2 \right].
\label{5}
\end{align}

A bound on uncertainty $\overline{j}^2/{\sigma_j}^2$ is obtained from the inequality~\eqref{PV} by constructing the total entropy production as the logarithm of the ratio of probabilities for a trajectory in extended state space and its time reversal. Within this picture, it can easily be shown that the mean entropy production per time step is 
\begin{eqnarray}
\bar{s}_{\text{tot}} = (1-f)(p'-q') \ln \left(\frac{p'}{q'}\right),
\label{RT2}
\end{eqnarray}leading to a ``naive'' PV bound  using this $\bar{s}_{\text{tot}}$ calculated in the extended state space.  Note that \eqref{RT2} has no $p$-dependence as here the entropic contributions associated with tumbles cancel out on average (because in the extended state space the number of changes from positive to negative preferred direction is asymptotically equal to that from negative to positive).\footnote{If instead the preferred direction is treated as an odd-parity variable, with signs flipped in the reversed trajectory,  one obtains a $p$-dependent mean entropy production rate  which, however, does not always bound the uncertainty (for related discussion on entropy production in active matter, see \cite{shankar2018}). } As we shall see the PV bound turns out to be very loose in many regions of parameter space (in particular, for  intermediate $p$ values); indeed as $p' \rightarrow 1$, $\bar{s}_{\text{tot}} \rightarrow \infty$.

Mathematically  the time-integrated current is described by a renewal-reward process (a type of cumulative process)\cite{RRTmain}, where tumbles are renewal events, and current increments from each run are rewards. In the next two sections, we show that this framework allows us to construct a general run-and-tumble bound on the uncertainty  which, significantly, also applies to non-geometric run lengths and is often considerably tighter than the PV bound.
\section{Renewal-reward theory (RRT)}
\begin{figure}
\onefigure[scale =0.1]{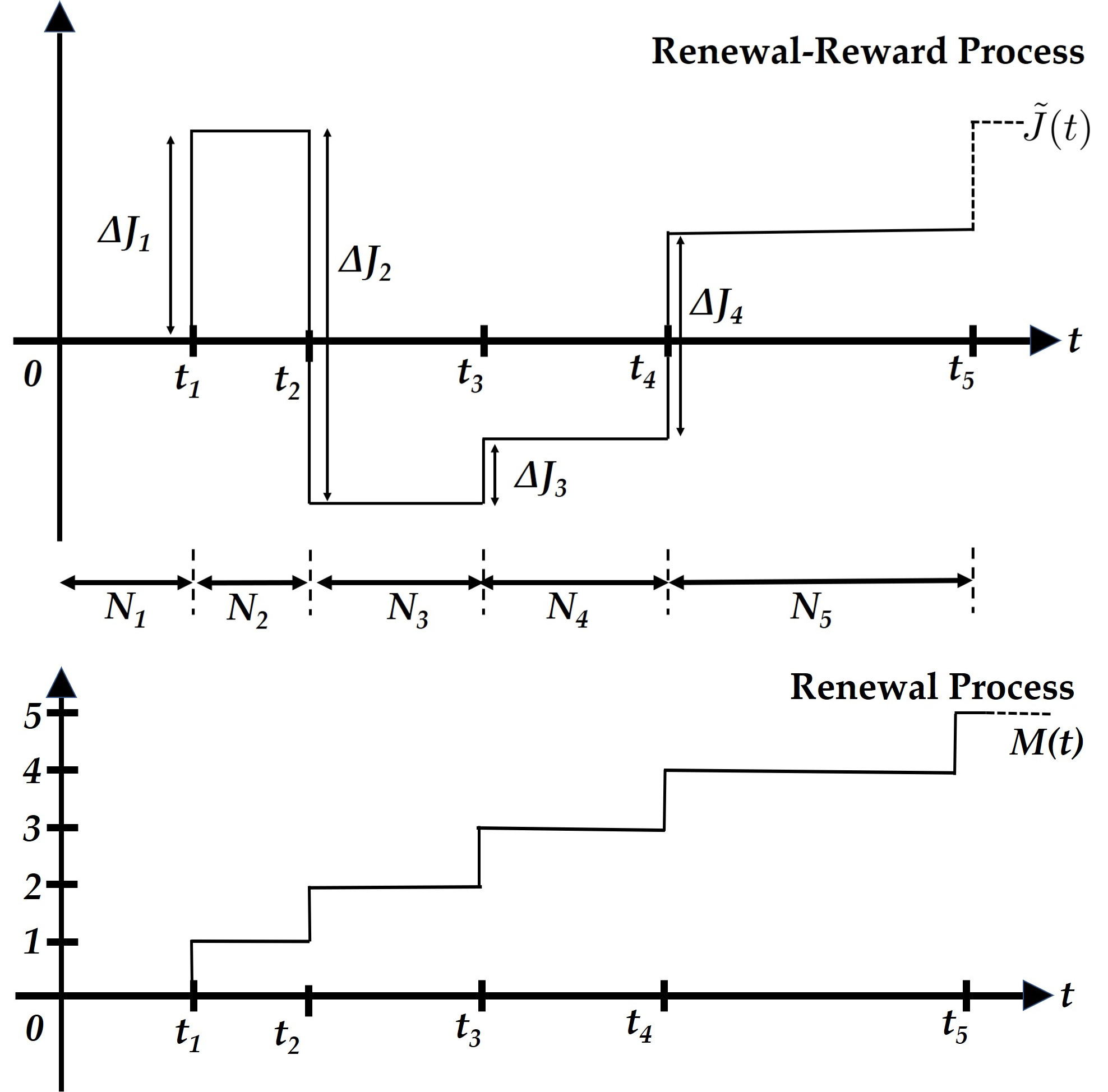}
\caption{Sample realisation of renewal process and corresponding renewal-reward process with tumbles (renewals) at $0, t_1,t_2, \ldots$ and current increments (rewards) $\Delta J_1, \Delta J_2, \Delta J_3\ldots $.}
\label{fig2}
\end{figure}
To build our general framework,  we start by considering the statistics of the tumble events. Let $M(t)$ be the total number of tumbles during the run-and-tumble process from time step 1 up to time step $t$. We assume that the interoccurrence times between tumbles are non-negative, independent and identically distributed (IID) random variables $N_i$, $i\geq1$, drawn from a discrete probability distribution.  For the toy model of the previous section the $N_i$'s are geometrically distributed but, in principle, we can take any distribution with finite mean ($0 < E[N_i] < \infty$). Under these assumptions,  $M(t)$ represents a renewal process (see lower part of Fig.~\ref{fig2}) where $M(t) = \text{max}\{m: \sum_{j =1}^m N_j\leq t\}$.

We now turn our attention to the current $J(t)$ which consists of the sum of current increments $\Delta J_i$ from completed runs and $\Delta J_{\text{F}}$ from the residual (uncompleted) run between the last tumble and time step $t$:
\begin{eqnarray}
J(t) = \sum_{i =1}^{M(t)} \Delta J_i  + \Delta J_{\text{F}}.
\label{6}
\end{eqnarray} In the case of geometric run lengths $\Delta J_{\text{F}}$ is from the same distribution as  the $\Delta J_i$'s but in general, this will not be true. However, at least in the case where all moments  of $N$ are finite, we expect that $J(t)$ is well approximated in the long-time limit by 
\begin{eqnarray}
\tilde{J}(t) = \sum_{i =1}^{M(t)} \Delta J_i. 
\label{7}
\end{eqnarray}If the current increments are independent of one another (as in the random walk model), then $\tilde{J}(t)$ is a so-called renewal-reward process where we can consider the current increment $\Delta J_i$ to be a terminal
``reward" added at the end of the $i$th run (see upper part of Fig.~\ref{fig2}). Note that each $\Delta J_i$ is a random variable which can be negative and depends on the direction and time of only the last tumble (not the previous history).

In particular, we are interested in processes where the current increment $\Delta J_i$ can be factorised as the product of independent random variables $X_i$ (set in the tumble) and $R_i$ (depending only on the run length $n_i-1$):
\begin{eqnarray}
\Delta J_i = X_i R_{i}. 
\label{8}
\end{eqnarray}For example, in our toy model,  $X_i$ takes values $\pm1$ and (for a run of length $n_i-1$) we have $R_i =2 \tilde{R	_i}-(n_i-1)$ where the number of forward steps $\tilde{R_i}$  has a binomial distribution $\text{B}(n_i-1,p')$. We are chiefly interested in the long-time behaviour of $\tilde{J}(t)$ which is obviously related to the moments of the run length and current increment distribution. For brevity, in what follows we drop the `$i$' subscript in notation for the moments of IID random variables. We define $\mu_k = E[N^k],~\lambda_k = E[\Delta J^k]$ and $c_{lk} = E[N^l \Delta J^k]$ where $k,l\geq1$ and assume these expectations are finite. Using standard asymptotic results in renewal-reward theory, we can write the long-time mean of $\tilde{J}(t)$ as:
 \begin{align}
 \lim_{t \rightarrow \infty} \frac{E[\tilde{J}(t)]}{t} &=  \frac{\lambda_1}{\mu_1} = \frac{E[X] E[R]}{E[N]},
 \label{meancurr}
 \end{align}where we utilise the fact that $X_i$ and $R_i$ are independent for all $i$. This is the mathematical expression of the intuition that the time-averaged mean current is asymptotically given by the expected current accumulated in one run divided by the expected time between tumbles.

Renewal-reward theory also provides an expression for the long-time scaled variance  of $\tilde{J}(t)$ as~\cite{RRTmain1, r4} 
  \begin{eqnarray}
 \lim_{t \rightarrow \infty} \frac{\text{Var} [\tilde{J} (t)]}{t} &=&  {\mu_1}^{-3} \mu_2  {\lambda_1}^2 - 2 {\mu_1}^{-2} c_{11} \lambda_1  \nonumber\\
&&\phantom{=} + {\mu_1}^{-1} \lambda_2,
 \label{variance}
\end{eqnarray} 
where in our set-up, $\mu_2=E[N^2],~\lambda_2=E[X^2] E[R^2]$ and $c_{11} = E[X] E[RN]$.

Given the assumption that the long-time statistics of $\tilde{J}(t)$ and $J(t)$ are the same, renewal-reward theory provides a natural structure to obtain the exact asymptotic uncertainty in terms of moments of the underlying random variables $R$, $X$, and $N$. In the next section, we will see that useful  bounds on the uncertainty can still be obtained without knowledge of the distribution of $R$. The key step is to use the result~\eqref{variance} to relate the variance of the current to that of a simpler Markovian process (associated with the tumbles) with known entropic bounds.

\section{Uncertainty bounds}
In this section, we outline the procedure to derive an entropic bound on the stochastic current fluctuations for the general class of run-and-tumble type processes. We now assume,
\begin{eqnarray}
E[R|N=n] =  \bar{r}(n-1),\quad\text{Var}[R|N = n] = {\sigma_r}^2 (n-1),
\label{11}
\end{eqnarray}where $\bar{r}$ and ${\sigma_r}^2$ are constants depending on the details of the run process. The scaling in \eqref{11} is clearly exact for random walks with IID step lengths such as the toy model above. Identifying $  {\sigma_j}^2 = \lim_{t \rightarrow \infty} \text{Var} [\tilde{J} (t)]/t$, $\bar{j} = \lim_{t \rightarrow \infty} E[\tilde{J}(t)]/t$, and rewriting \eqref{meancurr} and \eqref{variance} in terms of $\bar{r}$, ${\sigma_r}^2$, $\bar{N} = E[N]$, ${\sigma_N}^2 = \text{Var}[N]$, $\bar{X} = E[X]$, and ${\sigma_X}^2 = \text{Var}[X]$  gives
\begin{align}
\bar{j} &= \frac{\bar{X} \bar{r} (\bar{N}-1)}{\bar{N}},   \label{12} \\
{\sigma_j}^2 &= {\sigma_X}^2 \left[\frac{\bar{r}^2 (\bar{N}-1)^2}{\bar{N}} \right]  + {\sigma_r}^2 \left[\frac{\bar{X}^2 (\bar{N}-1)}{\bar{N}} \right] \nonumber\\
&\phantom{=}+ {\sigma_N}^2  \left[ \frac{\bar{X}^2 \bar{r}^2}{\bar{N}^3} \right] + {\sigma_X}^2 {\sigma_r}^2 \left[ \frac{(\bar{N}-1)}{\bar{N}} \right] \nonumber\\
&\phantom{=}+ {\sigma_X}^2 {\sigma_N}^2 \frac{\bar{r}^2}{\bar{N}} \label{13}.
\end{align}
Crucially, we note that all terms in \eqref{13} are positive which implies that by considering only some subset of them we can get a bound on ${\sigma_j}^2$. In particular, we have
\begin{eqnarray}
{\sigma_j}^2 \geq \left[\frac{\bar{r}^2  \left((\bar{N}-1)^2 + {\sigma_N}^2\right)}{\bar{N}} \right] {\sigma_X}^2,
\label{varbound}
\end{eqnarray}
where the tightness of this bound obviously depends on the relative size of different contributions in eq.~\eqref{13}; we expect it to be well-suited for cases when the run lengths are long and tumble statistics dominate the variance. A direct uncertainty bound follows from eqs.~\eqref{12} and \eqref{varbound}:
 \begin{eqnarray}
 \frac{\bar{j}^2}{{\sigma_j}^2} \leq \left[ \frac{(\bar{N}-1)^2}{\bar{N}((\bar{N}-1)^2 + {\sigma_N}^2)}\right] \frac{\bar{X}^2}{{\sigma_X}^2}. 
 \label{14}
 \end{eqnarray}
 
 \begin{figure}
\onefigure[width = 0.98\linewidth]{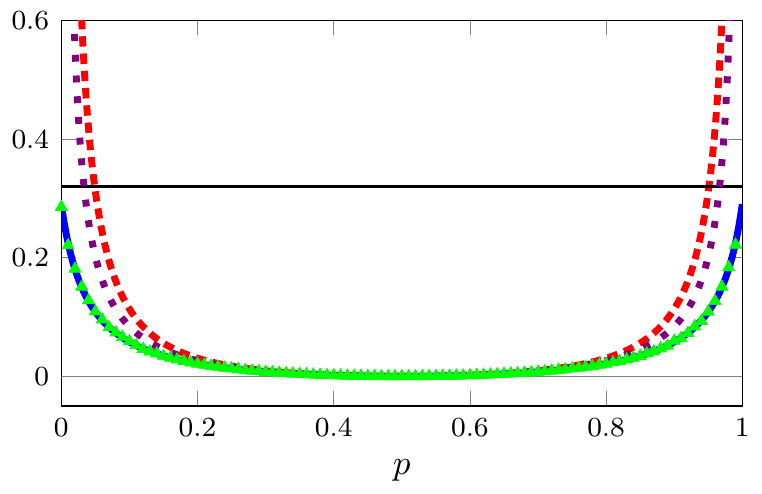}
\caption{RT bound [dashed red, \eqref{RTbound}], direct bound [dashed violet, \eqref{14}], PV bound [thin black, \eqref{PV}], and RRT prediction [solid blue, (\ref{12},\ref{13})] for geometrically distributed runs with $f = 0.1$, $p' = 0.75$. Green triangles show simulation results for $T =20000$ averaged over 10000 realisations. }
\label{fig3}
\end{figure}

The quantities in \eqref{14} have straightforward physical interpretations. However, to establish a connection with the standard entropic bounds of TURs, we now construct an auxiliary process by summing  IID random variables, $\mathsf{X}(M) = \sum_{i=1}^M X_i$; note that $\bar{X}^2$ and ${\sigma_X}^2$ are also the scaled cumulants of $\mathsf{X}(M)$. This process has discrete-time Markovian dynamics and assuming so-called ``microscopic reversibility'' (\textit{i.e}., if  $\text{Pr}(X_i = + x)$ is non-zero, then $\text{Pr}(X_i = -x)$ is also non-zero), we can trivially obtain its entropy production via the ratio of probabilities of state-space trajectories. Denoting the mean entropy production (per tumble step) as $\bar{s}_X$, the standard PV bound on the uncertainty of $X$ is 

\begin{eqnarray}
\frac{\bar{X}^2}{{\sigma_X}^2} \leq \frac{1}{2} \left(e^{\bar{s}_X} -1 \right).
\label{15}
\end{eqnarray}
 Combining eqs.~\eqref{14} and \eqref{15} leads us to the inequality
\begin{eqnarray}
 \frac{\bar{j}^2}{{\sigma_j}^2} \leq \frac{(\bar{N}-1)^2}{2 \bar{N} \left[ (\bar{N}-1)^2 + {\sigma_N}^2 \right]}  \left(e^{\bar{s}_X} -1 \right),
 \label{RTbound}
 \end{eqnarray}
 which we dub the ``RT bound'' in allusion to its structure  as a product of a prefactor depending on the run-length statistics and a term involving the entropy of the tumbles. Equation~\eqref{RTbound} can be expressed in the form shown
in \eqref{introRT}.

This bound is arguably more useful than the PV bound in situations where the microscopic dynamics of the run process is not readily accessible. [A weaker bound can also be obtained when the variance of the run lengths is not known, using only their mean.] In the next sections, we test our new RT bound for Markovian and non-Markovian run-and-tumble processes, and compare its tightness with the PV bound.

\section{Geometrically distributed runs}
\begin{figure}
\onefigure[width =1.001 \linewidth]{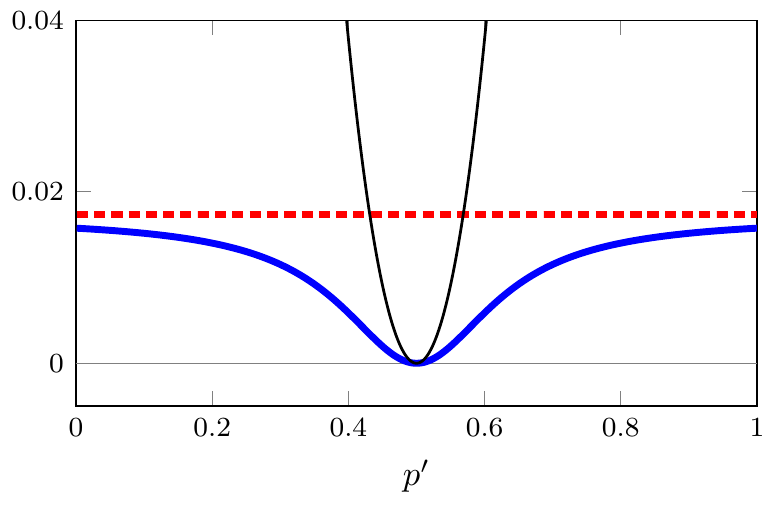}
\caption{Same as Fig.~\ref{fig3} but for fixed $p = 0.75$ and varying $p'$.}
\label{fig4}
\end{figure}
\begin{figure}
\onefigure[scale = 0.51]{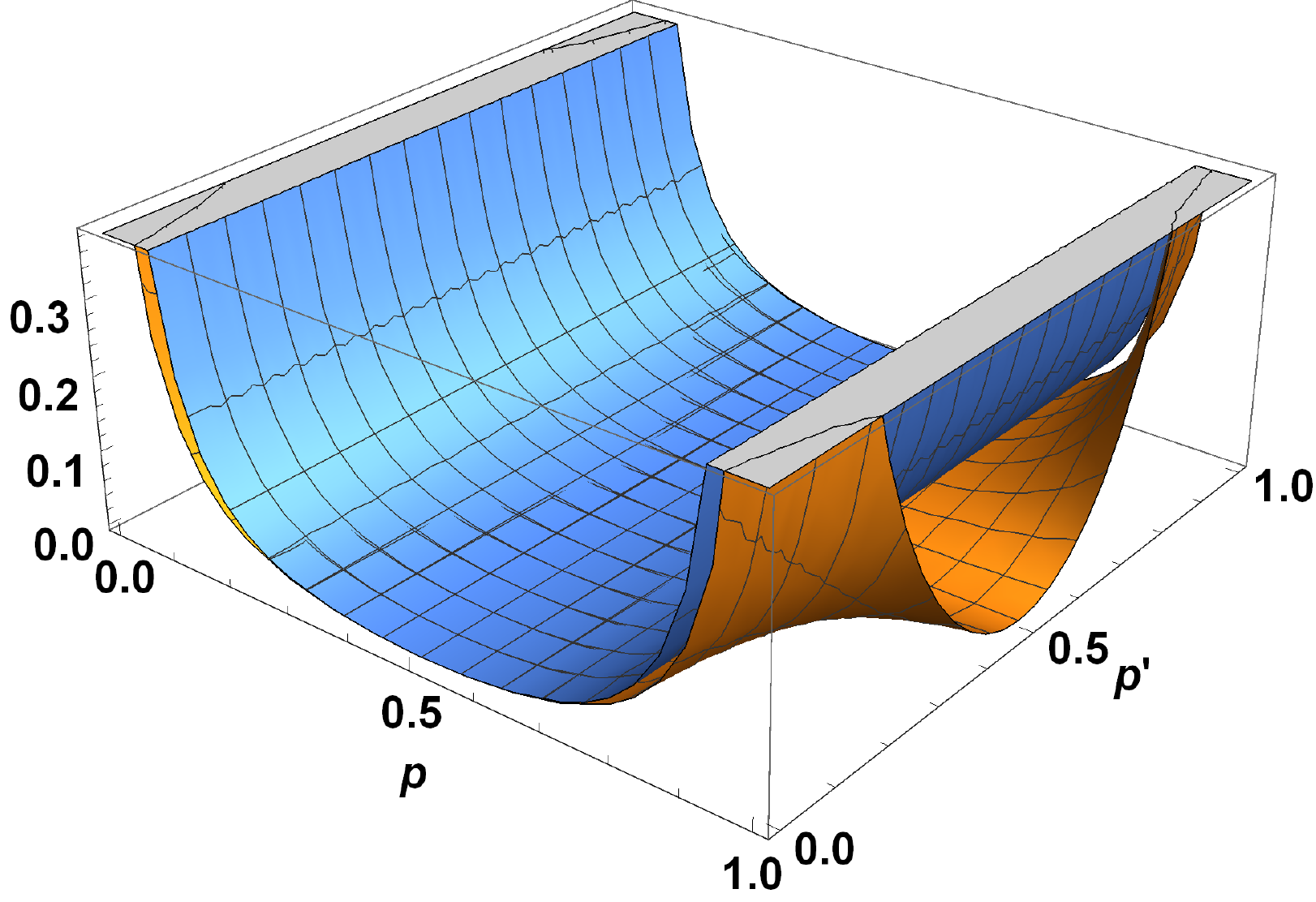}
\caption{RT bound (blue) and theoretical RRT uncertainty (orange)  as a function of $p$ and $p'$  for geometric runs with $f = 0.25$.}
\label{fig5}
\end{figure}To demonstrate the central result \eqref{RTbound}, we now consider tumbles of the specific form
\begin{eqnarray}
X_i =
\left\{
	\begin{array}{ll}
		1  & \mbox{with probability}~p, \\
		-1 & \mbox{with probability}~ q = 1-p,
	\end{array}
\right.
\label{17}
\end{eqnarray}
such that the mean auxiliary-entropy production rate is 
\begin{eqnarray}
\bar{s}_X = (p-q)\ln\left( \frac{p}{q} \right).
\end{eqnarray}
We first analyse our toy model with geometrically distributed run lengths. Here,  $\text{Pr}(N =n) =  f(1-f)^{n-1}$ where $f$ is the probability of tumbling and $n = 1,2,3,\ldots$. For the RT bound, we need only the first two cumulants: $\bar{N} = {1}/{f}$ and ${\sigma_N}^2   = (1-f)/f^2$. Hence, \eqref{RTbound} reduces to 
\begin{eqnarray}
 \frac{\bar{j}^2}{{\sigma_j}^2} \leq \frac{f\left(1-f\right)}{2\left(2-f\right)}  \left[\left(\frac{p}{q} \right)^{p-q} -1 \right].
 \label{GeomRT}
 \end{eqnarray}

We are chiefly interested in the behaviour of this bound for relatively long mean run lengths (corresponding to small $f$);  in Fig.~\ref{fig3} we plot the bound as a function of $p$ for $f = 0.1$ (mean run length $\bar{N}-1$ = 9) and compare it to the exact asymptotic uncertainty from renewal-reward theory and to Monte Carlo simulation. The numerics agree with the exact asymptotics as expected and they clearly obey the inequality~\eqref{GeomRT}. We observe that this RT bound is close to bound \eqref{14} and relatively tight for intermediate $p$ values but becomes loose as $p$ approaches 0 or 1 (cases where the model resembles a lazy random walker). In contrast, we also show the $p$-independent PV bound, obtained by using eq.~\eqref{RT2} in~\eqref{PV}, is only tight when $p$ approaches 0 or 1. For completeness, we also compare in Fig.~\ref{fig4} the RT and PV bounds at fixed $p$ as a function of $p'$ (although, we anticipate our results to be most useful when $p'$ is unknown); again we see that the bounds are tight in complementary regions. Obviously, the RT bound is a less informative constraint for larger $f$ (shorter run length) since the inequality \eqref{varbound} becomes looser. However, even for $f =0.25$ (mean run length 3) we see from the three-dimensional plot in Fig.~\ref{fig5} that the bound is reasonably tight in much of the parameter space. We now extend our analysis to non-geometric runs as may be relevant in applications.
\section{Other run distributions} The geometric run length distribution corresponds to a  Markovian process (on the state space of position and preferred direction) since the probability of tumbling is independent of the time elapsed since the last tumble. This may not be a good approximation in many real-life situations, \textit{e.g}., if energy needs to build up via a sequence of internal chemical reactions before a tumble can take place.  Also, there is much theoretical interest in fluctuations in non-Markovian processes so it is significant that our RT bound can be applied to arbitrary discrete run distributions with support on strictly positive integers. [Recall that for $n=1$, the corresponding run length is zero.] Although entropy production is in general difficult to compute for non-Markovian dynamics, the trajectory reversal argument for our class of models (where the tumbles form a semi-Markov process with ``time-direction independence"\cite{r6}) suggests that the analogue of eq.~\eqref{RT2} is 
\begin{eqnarray}
\bar{s}_{\text{tot}} = \left(\frac{\bar{N}-1}{\bar{N}}\right)(p'-q') \ln \left(\frac{p'}{q'}\right)
\label{RT3}
\end{eqnarray}
so we can also test the PV bound \eqref{PV} in this case.
For ease of comparison, we choose parameters for the plots in this section so that our non-geometric distributions have the same $\bar{N}$ as the geometric case with $f = 0.1$.
 \begin{figure}
\onefigure[width = \linewidth]{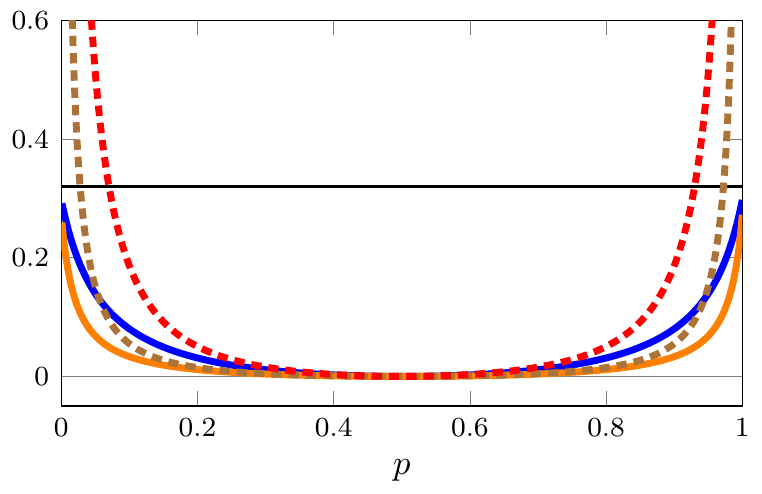}
\caption{RT bound (dashed),  RRT prediction (thick solid) and PV bound (thin black) for (i) negative binomial distribution with $f = 0.3$ and $k=3$ (red and blue) and  (ii) log-series distribution with $f' = 0.0269$ (brown and orange). }
 \label{fig6}
\end{figure}

A natural starting point to model a sequence of intermediate steps is to consider the negative binomial distribution corresponding to the sum of $k$ geometric random variables.  Here, $\text{Pr}(N=n) = \binom{n-1}{k-1} f^k (1-f)^{n-k}$ where $n = k, k+1, k+2, \ldots $ . When $k=1$, $\text{Pr}(N=n)$ reduces to the geometric distribution.  The mean and variance of $N$ are $\bar{N} = {k}/{f}$ and ${\sigma_N}^2   = k(1-f)/f^2$.  Hence, \eqref{RTbound} becomes 
\begin{eqnarray}
 \frac{\bar{j}^2}{{\sigma_j}^2} \leq \frac{f (k-f)^2}{2 k \left(f^2-3 f k+k^2+k\right)} \left[\left(\frac{p}{q} \right)^{p-q} -1 \right].
 \label{negbioRT}
 \end{eqnarray}
  In Fig.~\ref{fig6}, we show the RT bound and exact RRT results as a function of $p$ for $f = 0.3$ (mean run length 9) and observe features similar to the geometric case -- in particular, the RT bound is tighter than the PV bound for intermediate $p$. [We conjecture that validity of the PV bound is connected to time-direction independence.]

We now repeat our analysis for a more ``exotic" distribution, namely the log-series distribution defined as $\text{Pr}(N=n) = -(1-f')^n/(n \ln f')$ with parameter $f'$. The required moments are written as $\bar{N} = {(f'-1)}/{(f' \ln f')}$ and ${\sigma_N}^2   = (f'-1) (\ln f' -f'+1)/(f'^2 (\ln f')^2)$, and so in this case, the RT bound takes the form 
 \begin{eqnarray}
 \frac{\bar{j}^2}{{\sigma_j}^2} &\leq &  \frac{f' (1-f'+f' \ln f')^2}{2 (1-f') \left(1-3f'+2 f'^2-f'^2 \ln f' \right)} \nonumber \\
&& \phantom{=}\times \left[\left(\frac{p}{q} \right)^{p-q} -1 \right].
 \label{20}
 \end{eqnarray}
Figure~\ref{fig6} confirms that again our bound yields a useful constraint. We have also checked that a similar picture is obtained for the zero-truncated Poisson process. For an extreme delta-like distribution (fixed run length $\bar{N}$), the RT bound reduces to the form of \eqref{PV} with $\Delta t = \bar{N}$ which is intuitively reasonable since in this case the process resembles a random walk with longer time step and increased variance due to the contribution of ${\sigma_r}^2$ terms in \eqref{13}.
\begin{figure}
\centering 
\includegraphics[width = \linewidth]{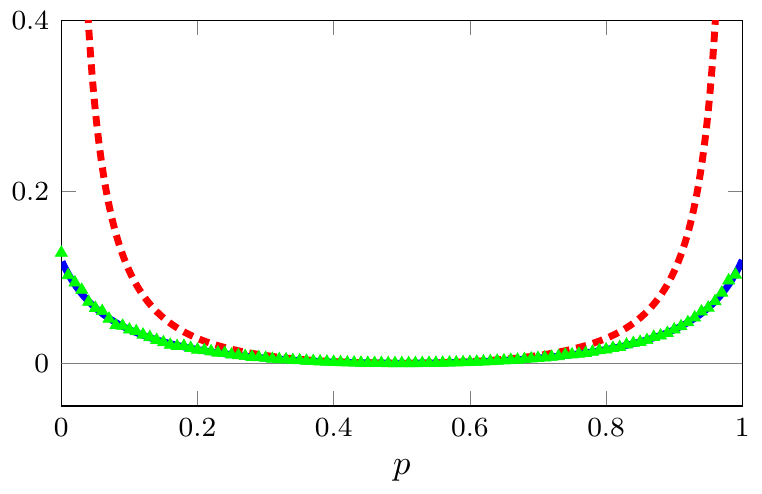}
\caption{RT bound for ASEP with 5 particles on a 10-site ring. Particles hop (subject to the exclusion constraint) in the preferred direction with rate 0.75 and in the opposite direction with rate 0.5. The preferred direction itself is stochastically reset (clockwise with probability $p$) and the generalised run is exponential with $\bar{N}-1$ = 9. Key as in Fig.~\ref{fig3} with simulation results for $T= 20000$ averaged over 2000 realisations.}
\label{fig7}
\end{figure}
\section{Discussion}
In this letter, we have demonstrated a simple run-and-tumble (RT) bound for single-particle models in discrete time. Since the renewal-reward framework on which the derivation is based also holds in a continuous-time setting~\cite{r4} and the auxiliary process is discrete-time by construction, the same bound (\ref{RTbound}) should apply to continuous-time models as well. Significantly, we also anticipate it is applicable to many-particle systems where the preferred direction is stochastically reset at random times (which can be construed as a ``collective tumble" for all the particles). The assumptions required for our  asymptotic bound are that current increments are IID with the form~\eqref{8}, and that the mean and variance of the ``generalised run" process between resets scale as in~\eqref{11}. In fact, for many-particle systems that scaling may not hold exactly but it is generically true in the large $n$ limit, so even in this case, the RT bound is expected to be useful when the collective tumbles are infrequent.

As an example of an interacting particle system, we show results in Fig.~\ref{fig7} for a paradigmatic model --- the  
asymmetric simple exclusion process (ASEP)\cite{asep2} on a ring. Here the usual continuous-time exclusion dynamics is punctuated by a random resetting of the preferred direction (clockwise/anti-clockwise);  in direct analogy to our discrete-time single-particle models, we assume that this reset event takes one unit of time. A theoretical expression for the exact uncertainty can be easily derived using  eqs.~\eqref{12} and \eqref{13} together with known results for $\bar{r}$ and ${\sigma_r}^2$~\cite{asep1, asep1A}, and is verified by simulation. Fig.~\ref{fig7} confirms that this uncertainty indeed obeys the RT bound.

To conclude, we suggest that  although a tighter bound could be obtained including more terms in \eqref{13}, the power of our RT bound is that knowledge of only the mean and variance of runs, and the mean entropy production rate associated with tumbles is enough to infer constraints on the current fluctuations. We emphasise that this bound is independent of the parameters of the underlying run process (\textit{e.g.,}  hop rates in the ASEP). With more information about the microscopic dynamics, one could potentially derive other bounds (in the spirit of~\cite{i8,i9}) using the large deviation formalism~\cite{LDTRRT}. It would also be interesting to investigate the link between multivariate renewal-reward theory~\cite{mRRT} and proposed multidimensional uncertainty relations~\cite{mTUR}. Finally, the applicability of the bound to models exhibiting dynamical phase transitions~\cite{r1} remains to be explored; recent works~\cite{PT, PT1, PT2} suggest that this is relevant for various run-and-tumble applications.
\acknowledgments
The authors thank Patrick Pietzonka and Robert L. Jack for useful discussions. RJH gratefully acknowledges the support of the London Mathematical Laboratory in the form of an External Fellowship.
\renewcommand{\theequation}{A-\arabic{equation}}
  % redefine the command that creates the equation no.
  \setcounter{equation}{0}  % reset counter 
%  \section*{APPENDIX}
\section{Appendix: reset framework}
As alluded to in the main text, the toy run-and-tumble model can be understood as an asymmetric random walk with intermittent \textit{resets} of preferred direction. Reset processes are of considerable topical interest~\cite{a1,a2,a3,a4,a5,a6,a7} and there is now a well-understood framework allowing the calculation of large deviations and the identification of phase transitions~\cite{r1}.  To allow for correlation between the current in the run and the direction of the preceding tumble, we here adapt the approach of~\cite{r1} by considering the current generating function $W(s,n)$ for a combined tumble-and-run event of duration $n$ steps. In principle, the generating function for the trajectory current $J(t)$ can then be found by summing over all possible combinations of tumble-and-runs with total duration $t$. In practice, it is easier to relax this constraint by switching to Laplace space; the \textit{z}-transform of $W(s,n)$ is 
\begin{eqnarray}
\tilde{W}(s,z) = \sum_{n=1}^{\infty} W(s,n) z^{-n},
\label{A-2}\end{eqnarray} 
where $z$ is the conjugate parameter to $n$. Since any number of tumble-and-runs is now allowed, the \textit{z}-transformed generating function for $J(t)$ is a geometric sum with ratio $\tilde{W}(s,z)$. The long-time behaviour is controlled by $z^{*}$, the largest real value of $z$ for which the sum diverges. [In the absence of phase transitions, we simply set $\tilde{W}(s,z)=1$.] In particular, the desired SCGF $\phi(s)$ is given by $\ln z^{*}$.

Applying this method for geometrically-distributed runs with parameter $f$ gives
\begin{align}
W(s,n) &= f (1-f)^{n-1} [p (p'e^{+s} +q' e^{-s})^{n-1}\nonumber\\
& \phantom{}+ q (p' e^{-s} +q' e^{+s} )^{n-1}],
\label{A-3}
\end{align}
and hence
\begin{align}
\phi(s) &=\ln \frac{1}{2} \Bigg( f + 2(1-f)\cosh (s) + \Big\{2(1-f)(p'-q') \nonumber\\
&\phantom{=}\times \left[2 f (p-q)\sinh (s) + (1-f)(p'-q')\cosh(2s)\right]\nonumber\\
& \phantom{=} -(f^2 - 4f +2) + 8(1-f)^2 p' q' \Big\}^{1/2} \Bigg),
\label{A-4}
\end{align} where the positive root is taken to ensure $\phi(0) = 0$. For non-geometric run distributions it may be more difficult to obtain the full SCGF via this resetting approach; however, the RRT framework of the main text still provides an efficient method to retrieve the first two cumulants.

\bibliographystyle{eplbib}
\bibliography{eplref1}

\end{document}